\documentclass[preprint]{vgtc}

\usepackage{amsmath}
\usepackage{dblfloatfix}   % allows figure* to float to bottom as well as top

\graphicspath{{figs/}{figures/}{pictures/}{images/}{../}}

\abstract{
The application of machine learning to molecular property prediction has become
increasingly prevalent in drug discovery, yet most models operate as black boxes,
returning a prediction without revealing which structural features drive it.
MolExplain addresses this gap by combining property prediction with sub-structure
level visual explainability in an interactive web interface. The system featurizes
molecules as Morgan fingerprints, classifies them using a trained XGBoost model,
and applies SHAP attribution to produce a smooth heatmap overlay indicating which
regions of the molecule contribute for or against the predicted property. Applied
to cyclic peptide membrane permeability, the tool's attribution independently
recovers the known role of backbone N-methylation in improving passive membrane
diffusion, consistent with established chemistry. While demonstrated
on cyclic peptides, the framework is designed to generalize to other molecular
properties, positioning MolExplain as a platform for interactive,
explainability-driven molecular design.
}

\vgtccategory{Research}
\onlineid{1573}

\begin{document}

\title{MolExplain: An Interactive Tool for Explainable Molecular Property Prediction}

\author{Pirm Dhararag \and Dylan Cashman}

\author{\authororcid{Pirm Dhararag}{0009-0007-7060-3199}\thanks{e-mail: pirmdhararag@brandeis.edu}\\ %
        \scriptsize Brandeis University %
\and \authororcid{Dylan Cashman}{0000-0003-4853-5701}\thanks{e-mail: dylancashman@brandeis.edu}\\ %
     \parbox{1.4in}{\scriptsize \centering Brandeis University}}

% The paper headers
\markboth{Dhararag and Cashman: MolExplain}%
{Dhararag \MakeLowercase{\textit{et al.}}: MolExplain}

\maketitle

%-------------------------------------------------------------------
\section{Introduction}

Machine learning has become an increasingly valuable tool across the drug discovery
pipeline, enabling rapid prediction of molecular properties such as solubility,
toxicity, binding affinity, and membrane permeability directly from chemical
structure~\cite{Dara2022}. While these models have grown more accurate, a persistent
limitation remains: most operate as black boxes~\cite{JimenezLuna2020,Gangwal2026}.
They return a prediction without revealing which structural features drive it.
Stokes et al., for example, used a deep neural network to identify halicin as a
novel antibiotic candidate from over 100 million molecules, yet because the model
was agnostic to the mechanism of action underlying the prediction, understanding why
halicin was active required entirely separate experimental
investigation~\cite{Stokes2020}. For a medicinal chemist attempting rational design,
a prediction without structural explanation leaves no principled basis for deciding
where to intervene, and this gap between prediction and insight is a fundamental
challenge across property prediction tasks in drug discovery.

Cyclic peptide membrane permeability represents a compelling case where this
limitation matters. Compared to their linear counterparts, cyclic peptides exhibit
enhanced structural rigidity, improved receptor selectivity, and greater resistance
to proteolytic degradation, properties that have contributed to a steady increase
in clinically approved peptide-based therapeutics over the past two decades.
Despite these advantages, a central challenge in cyclic peptide drug development
remains their ability to passively cross cellular membranes. Membrane permeability
is governed by physicochemical factors such as lipophilicity, hydrogen bonding
capacity, conformational flexibility, and polar surface area~\cite{Rezai2006,Linker2023}.
Chemical strategies such as backbone N-methylation have been explored to improve
permeability by reducing hydrogen bond donors~\cite{Rader2018,Chatterjee2012}, but these modifications increase
synthetic complexity and do not guarantee success. Recent predictive models including
transformer-based architectures such as PeptideCLM~\cite{Feller2025} and multimodal
deep learning models such as Multi\_CycGT~\cite{Cao2024} have demonstrated competitive
performance on this task, yet neither exposes structural reasoning to the user,
leaving them without actionable insight into which parts of the molecule to
investigate further.

MolExplain addresses this gap by combining property prediction with sub-structure
level visual explainability in an interactive web interface. The system accepts a
molecular structure as input, either drawn using a structure editor or provided as
a SMILES string, and returns both a predicted property probability and a smooth
heatmap indicating which regions of the molecule contribute for or against the
predicted property. While cyclic peptide membrane permeability serves as the
demonstrating use case in this work, the framework is designed to generalize to
other molecular properties with the addition of new models and labeled datasets.
The design is organized around three core user tasks, described in Section~3.
This work makes the following contributions:
\begin{itemize}
  \item MolExplain, an interactive web tool integrating molecular property prediction
        with substructure-level visual attribution in a single interface
  \item A use case demonstrating that Morgan fingerprint-based SHAP attribution
        independently recovers a known structural driver of cyclic peptide membrane
        permeability
  \item A system architecture designed for extensibility to additional molecular
        properties with minimal implementation overhead
\end{itemize}

% Figure 1: declared here so LaTeX queues it for top of page 2
\begin{figure*}[!t]
    \centering
    \includegraphics[width=\linewidth]{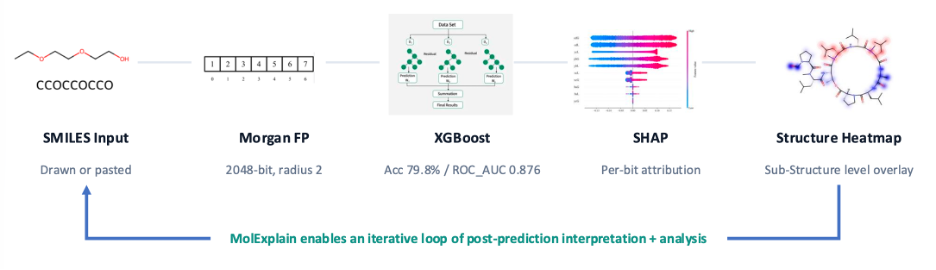}
    \caption{The MolExplain prediction pipeline. A SMILES string is converted to a
    2048-bit Morgan fingerprint (radius~2), which is passed to a trained XGBoost
    classifier (accuracy 79.8\%, ROC-AUC 0.876). SHAP TreeExplainer assigns
    per-bit attribution scores, which are mapped back to individual atoms via
    RDKit's bitInfo and smoothed into a sub-structure level heatmap overlay.
    The feedback arrow illustrates MolExplain's support for iterative
    post-prediction interpretation and structural modification.}
    \label{fig:pipeline}
\end{figure*}

%-------------------------------------------------------------------
\section{Related Work}

\subsection{Molecular Property Prediction}

Morgan fingerprints remain one of the most effective and widely used molecular
representations for property prediction tasks. Zhou and Skolnick demonstrated their
utility in virtual ligand screening, showing that circular substructure encodings
capture chemically meaningful patterns across diverse molecular
datasets~\cite{Zhou2024}. Their traceability through RDKit's bitInfo mapping, which
links each active fingerprint bit back to the atoms that generated it, makes them
particularly well-suited for substructure-level attribution workflows. While
gradient-based attribution methods have been developed for graph neural network
models in drug discovery contexts~\cite{JimenezLuna2020,Gangwal2026}, the direct
mapping between Morgan fingerprint bits and specific molecular substructures makes
SHAP attribution more tractable for the kind of chemist-interpretable heatmap
MolExplain requires. PeptideCLM and Multi\_CycGT are examples of deep learning-based
models applied to cyclic peptide membrane permeability
prediction~\cite{Feller2025,Cao2024}, but neither provides structural attribution to
the user. MolExplain is designed to address precisely this gap.

\subsection{Molecular Visualization}

The visualization of structure-activity relationships has a longer history in
cheminformatics. Guha introduced the concept of the ``glowing molecule,'' in which
individual atoms are colored by their contribution to a predicted property, providing
an intuitive visual link between molecular structure and model
output~\cite{Guha2013}. Building on this concept, Heberle et al.\ developed
XSMILES, an interactive visualization system that coordinates XAI attribution scores
across both SMILES token representations and 2D molecular
diagrams~\cite{Heberle2023}. MolExplain draws from both of these works in its
substructure-level heatmap design, while focusing on delivering this capability
through a self-contained web interface rather than a research visualization
environment.

\subsection{Visual Analytics for Interactive Machine Learning}

A broader trend in scientific AI has shifted explainability from a post-hoc
diagnostic mechanism to a core component of an interactive, human-in-the-loop
design process~\cite{Amershi2014}. Rather than treating model outputs as terminal
results to accept or reject, this paradigm positions domain experts as active
participants who apply their knowledge to form hypotheses, direct exploration, and
refine conclusions within the same interface.

This principle has produced a substantial body of work within the visualization
community itself, generally organized around the idea that visual analytics systems
should support users in understanding, diagnosing, and refining machine learning
models as part of a single iterative workflow. Spinner et al.\ formalize this as an
explicit pipeline, proposing a framework in which users cycle between understanding
a model's behavior, diagnosing its limitations using explainable AI methods, and
refining the model in response~\cite{Spinner2020}. Hohman et al.\ address a related
need at greater scale, developing Summit, a system that aggregates activation and
attribution information across an entire dataset so that a user can identify
recurring, class-level patterns rather than inspecting single-instance explanations
in isolation~\cite{Hohman2020}. Cashman et al.\ present a visual analytics system for iteratively
discovering neural network architectures, in which users alternate between varying
model configurations and inspecting the resulting performance to guide the next
modification~\cite{Cashman2020}, and Das et al.\ describe BEAMES, an interactive
system for steering and inspecting multiple regression models simultaneously,
letting users adjust feature weights and immediately observe the effect on model
fit~\cite{Das2019}.

Two systems illustrate this same human-in-the-loop principle in domains closer to
MolExplain's own. Hazarika et al.\ present NNVA, in which a neural network surrogate
model lets computational biologists adjust expensive simulation input parameters for
a yeast cell polarization model and immediately visualize the predicted outcome,
closely mirroring MolExplain's predict-modify-resubmit loop~\cite{Hazarika2020}. Lange et al.\ illustrate the same principle at the
dataset level, presenting a generative AI system that enables researchers to
progressively query and interpret complex biological datasets through a linked
visualization dashboard~\cite{Lange2025}. MolExplain is motivated by this same
principle applied to molecular property prediction: structural attribution is
not an add-on feature but the mechanism through which the chemist and the model
reason together.

% Figure 2: declared here (end of Related Work) so LaTeX queues it earlier,
% giving it the best chance of landing at the top of page 3
\begin{figure*}[!t]
    \centering
    \includegraphics[width=\linewidth]{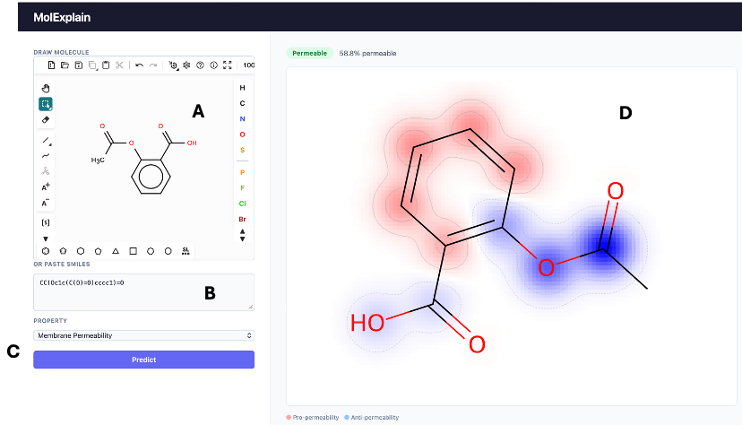}
    \caption{The MolExplain user interface on a small molecule. Panel A shows the interactive molecule
    sketcher. Panel B shows the SMILES text field, which stays in sync with the
    sketcher. Panel C contains the property selector and Predict button. Panel D
    displays the prediction result including the permeability label, confidence
    score, substructure-level heatmap, and color legend.}
    \label{fig:ui}
\end{figure*}

%-------------------------------------------------------------------
\section{Design Goals and Tasks}

MolExplain is organized around three user tasks that reflect the needs of a
medicinal chemist working with molecular property prediction. These tasks were
derived from the central limitation identified in Section~1: prediction without
structural explanation provides insufficient support for rational design decisions.
Each task below describes what the system must enable, why it matters, and what
design implications follow.

\subsection{T1: Property Prediction}

The first task is rapid property screening without wet-lab experiments. A medicinal
chemist working across a large library of candidate compounds cannot afford to
synthesize and test each one experimentally, and computational prediction enables
early-stage filtering that directs experimental resources toward the most promising
candidates. For MolExplain, this means the system must accept any valid molecular
structure and return a prediction quickly enough to support iterative use. Both
interactive drawing via a structure editor and direct SMILES input are supported so
the tool fits into existing workflows regardless of the user's preferred input mode.

\subsection{T2: Structural Attribution}

The second task is visual identification of which part of the molecule drives the
prediction. A probability score alone is not actionable. For instance, a chemist who
receives a prediction of 42\% permeable has no structural basis for deciding what to
change. T2 is what extends MolExplain beyond a standard classifier. The attribution
must be rendered at the substructure level as a visual overlay on the 2D molecular
diagram, so the chemist can connect the model's reasoning to the parts of the
molecule they can actually modify.

\subsection{T3: Iterative Exploration}

The third task is supporting a predict-modify-resubmit workflow. Rational molecular
design is not a single query but a loop. The chemist observes the attribution, forms
a hypothesis about which structural change would improve the property, makes that
modification, and resubmits. This system must support this loop without friction.
The input and output panels remain active after a prediction, and the Ketcher
sketcher stays synchronized with the SMILES field. Structural edits can be made
directly on the predicted molecule so the updated prediction reflects both the new
score and the new attribution pattern.

%-------------------------------------------------------------------
\section{Method}

\subsection{Dataset and Labeling}

The dataset used in this work is drawn from CycPeptMPDB, a curated database of
experimentally measured membrane permeability values for cyclic
peptides~\cite{Li2023}. Only entries with PAMPA measurements and valid SMILES
representations were retained. Permeability prediction is framed as a binary
classification task, with peptides assigned a positive label if their PAMPA value
is greater than or equal to $-6$ and a negative label otherwise. This threshold
reflects the boundary between permeable and non-permeable behavior in passive
membrane diffusion assays.

\subsection{Featurization and Model Training}

Each molecule is featurized as a Morgan fingerprint computed from its SMILES string
using RDKit, with a radius of 2 and a bit vector length of 2048. Morgan fingerprints
encode circular substructural environments around each atom, and the resulting binary
vector serves as the input to the classifier. An XGBoost model is trained on this
representation, achieving an accuracy of 79.8\% and a ROC-AUC of 0.876 on the
held-out test set. The full pipeline is illustrated in Figure~\ref{fig:pipeline}.

\subsection{SHAP Attribution}

The interactive visualization of MolExplain is achieved by bridging SHAP attribution
scores from the fingerprint level down to individual atoms, producing a
substructure-level explanation to the user. SHAP's TreeExplainer is applied to the
trained XGBoost model to compute a per-bit importance score for each of the 2048
fingerprint bits~\cite{Lundberg2017}. In our use case, a positive score indicates
that the bit pushed the prediction toward permeable, and a negative score indicates
the opposite. To translate these bit-level scores into per-atom contributions,
RDKit's bitInfo mapping is used. For each active fingerprint bit, bitInfo records
one or more (center atom, radius) pairs representing the circular substructural
environments that activated it. For each such environment, all atoms falling within
that radius are identified using RDKit's built-in function. The bit's SHAP score is
then divided equally among all atoms in that environment, and each atom accumulates
contributions across all bits it participates in.

\subsection{Heatmap Generation}

These per-atom scores are normalized to the range $[-1, 1]$ and passed to RDKit's
SimilarityMaps module to produce a continuous substructure-level heatmap overlaid
on the 2D molecular structure. The smoothing is what elevates the visualization
from discrete per-atom values to interpretable substructural regions, allowing the
user to identify functional groups and neighborhoods that drive the prediction
rather than individual atoms in isolation. Red regions indicate substructures
contributing toward permeability, while blue regions indicate substructures
contributing against it. The heatmap is rendered server-side as an SVG and returned
directly to the frontend. Because normalization is computed independently for each
prediction, heatmap intensity reflects the relative importance of substructures
within that single prediction rather than an absolute scale comparable across
separate predictions. It is intended to guide the user toward which regions of the
current molecule are worth inspecting or modifying next. Comparison across
iterations, such as tracking whether a modification improved the outcome, is
instead supported by the predicted probability itself, which is reported alongside
the heatmap at each step.

% Figure 3: declared here (end of \S4.4) so LaTeX queues it for top of page 4
\begin{figure*}[!t]
    \centering
    \includegraphics[width=\linewidth]{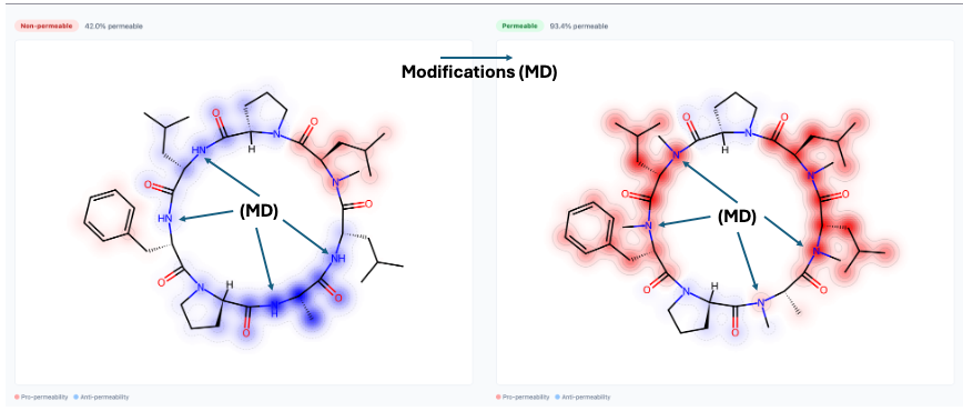}
    \caption{Use case walkthrough using MolExplain. Left panel shows the initial
    prediction of 42\% permeable, with unmethylated backbone nitrogen groups
    highlighted in blue as anti-permeability contributors. Right panel shows the
    result after iterative N-methylation of all backbone nitrogens, with the
    predicted permeability rising to 93.4\% and the modified regions now
    appearing red.}
    \label{fig:usecase}
\end{figure*}

%-------------------------------------------------------------------
\section{System Design}

\subsection{Architecture}

MolExplain is built as a two-component system: a React frontend served via Vite
and a Python backend built on FastAPI\footnote{Code is included via OSF Github integration at \url{https://osf.io/xbncm/overview?view_only=3f44172bb8ae45b9a4505f07a8423dc5}. Code will be open-sourced upon acceptance.}. All machine learning computation and
chemistry rendering are handled server-side, while the frontend is responsible
solely for user interaction and result display.

Three architectural choices directly reflect the design goals established in
Section~3. Morgan fingerprints were chosen over graph neural network representations
on both transparency and scale grounds: expert-defined fingerprints offer
transparency and lower computational cost for small-to-medium
datasets~\cite{Gangwal2026}, and RDKit's bitInfo mapping provides a direct,
deterministic link from each active bit to the atoms that generated it, without the
post-hoc attribution step required by GNN approaches such as GNNExplainer or
gradient-based methods~\cite{JimenezLuna2020,Gangwal2026}. Should future datasets,
whether expanded cyclic peptide permeability data or datasets for new properties
added to the platform, reach a scale where a graph-based approach becomes more
appropriate, GNN representations may be a natural extension. XGBoost was chosen as
the classifier because tree-based models are well-suited to the sparse binary tabular
structure of Morgan fingerprint inputs~\cite{Chen2016}, and because SHAP's
TreeExplainer computes exact, model-consistent attributions for tree ensembles. This
is an explicit trade of peak predictive accuracy for the interpretability T2
requires. SHAP was preferred over LIME because LIME's perturbation-based local
approximations are inherently nondeterministic, producing inconsistent explanations
across runs~\cite{Gangwal2026,Ribeiro2016}.

\subsection{Frontend}

The frontend interface is organized as a two-panel layout, illustrated in
Figure~\ref{fig:ui}. The left panel contains the molecule input components. Panel A
is a Ketcher molecule sketcher~\cite{Karulin2011} that allows users to draw structures interactively
using a full suite of chemical drawing tools. Panel B is a SMILES text field
positioned below the sketcher, which accepts pasted or typed SMILES strings. The
two inputs are kept in sync: typing into the SMILES field updates the Ketcher canvas
in real time and vice versa, allowing users to work in whichever input mode they
prefer. Panel C contains the property selector dropdown and the Predict button, which
triggers the API call on explicit user action rather than on every structural change.
The right panel, Panel D, displays the prediction output: a labeled badge indicating
permeable or non-permeable, the predicted probability, the substructure-level
heatmap, and a color legend.

A web interface was chosen over a notebook environment because the Ketcher sketcher
makes the predict-modify-resubmit workflow of T3 natural: the user modifies the
structure directly on the canvas and resubmits in a single interaction. A
notebook-based implementation would require re-entering or manually editing SMILES
strings between iterations, introducing friction that would impede the iterative
design loop.

\subsection{Backend and API}

Communication between the frontend and backend occurs through three REST API
endpoints: a prediction endpoint that accepts a SMILES string and property selection
and returns a prediction result, a properties endpoint that exposes the list of
available molecular properties, and a health endpoint for server status.

Two implementation decisions are worth noting. First, the heatmap SVG is rendered
server-side using RDKit's SimilarityMaps module and returned as a raw SVG string,
which the frontend renders directly in the output panel. This keeps all chemistry
rendering logic in Python and avoids any need to replicate RDKit's drawing
capabilities in the browser. Second, incoming SMILES strings are validated using
Pydantic before the pipeline executes, ensuring that malformed inputs are rejected
with an informative error before reaching the model.

\subsection{Extensibility}

The backend uses a model registry pattern where each supported molecular property is
defined as a configuration entry pointing to a saved model file. Adding a new
property requires only a new entry in the registry and a corresponding model file,
with no changes to the API or frontend. This design allows MolExplain to scale to
additional molecular properties and datasets without architectural changes,
supporting the platform vision described in the conclusion.

%-------------------------------------------------------------------
\section{Use Case}

Consider a cyclic peptide from the held-out test set during model training. To
make this a motivating example, let's assume it binds to a clinically relevant
intracellular target but fails to realize its therapeutic potential due to poor
membrane permeability. When loading this peptide into MolExplain, the tool returns
a predicted permeability of 42\%, consistent with its experimentally measured PAMPA
value of $-6.12$, which falls just below the permeable threshold of $-6$. The
alignment between the predicted score and the true PAMPA value immediately
establishes confidence in the model's output and motivates the user to examine the
heatmap for structural insight.

Examining the heatmap panel, the user, who is fluent in biochemistry, observes that
several regions are highlighted in red, indicating substructures contributing toward
permeability. These include the methylated backbone nitrogen, the alkyl side chains,
and the benzene ring on the side chain, all of which are well-established lipophilic
features that favor membrane permeability. However, a particularly deep blue region
is spotted at one of the unmethylated backbone nitrogen groups, with several other
backbone NH positions also appearing in blue. These blue regions indicate regions
actively working against permeability, and the depth of the color at the unmethylated
nitrogen draws the user's attention as a candidate for targeted modification.

Guided by this attribution, the user applies N-methylation to the flagged backbone
nitrogen and resubmits the modified structure. After several iterations of this
process, resulting in full methylation of all backbone nitrogen groups, the final
prediction rises to 93.4\% permeable. As shown in Figure~\ref{fig:usecase}, the
previously blue backbone nitrogen regions now appear red in the updated heatmap,
confirming that the modifications directly addressed the substructure features the
model had identified as anti-permeability contributors. The iterative workflow,
predict, interpret, modify, and resubmit, demonstrates T1, T2, and T3 in practice.
The hypothesis driving each structural modification is formed through a combination
of the heatmap's attribution pattern and the user's domain expertise. This level of
interaction between the model and the user is only possible because of the
explainability layer applied on top of the prediction. Without it, the user would
have only a probability score and no structural basis for deciding where to
intervene.

Importantly, backbone N-methylation is a well-established chemical strategy for
improving passive membrane permeability in cyclic peptides, as it reduces the
number of hydrogen bond donors and lowers the energetic cost of membrane
insertion~\cite{Rader2018,Chatterjee2012}. The fact that MolExplain's attribution
independently flags unmethylated backbone nitrogens as the primary
anti-permeability contributors, without any domain knowledge encoded beyond the SHAP
scores, is consistent with established chemistry. This provides evidence that the
Morgan fingerprint representation combined with per-bit SHAP attribution captures
chemically relevant structural signals.

%-------------------------------------------------------------------
\section{Discussion}

% Figure 4: zoomed heatmap, single-column — placed at top of Discussion column
\begin{figure}[!t]
    \centering
    \includegraphics[width=\columnwidth]{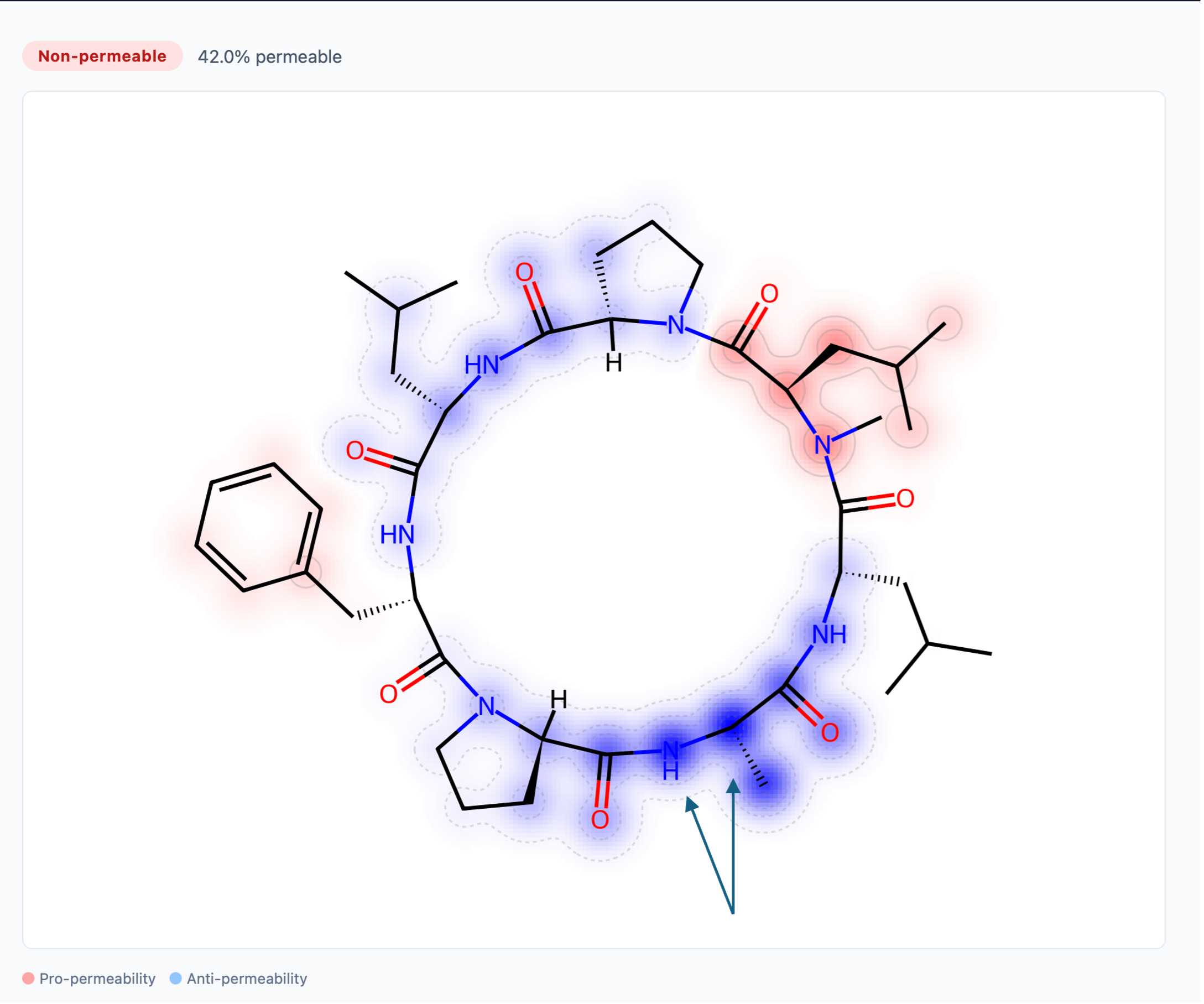}
    \caption{A zoomed heatmap of the test peptide highlighting a known limitation
    of fingerprint-based attribution. Arrows point to a methyl group adjacent to
    an unmethylated backbone nitrogen, both appearing in deep blue with nearly
    identical intensity. The methyl group's anti-permeability attribution likely
    arises not from its own chemical character but from its proximity to the NH
    group within shared Morgan fingerprint circular environments. A
    biochemistry-literate user would recognize the NH as the true driver in this
    region, but the overlap illustrates an inherent interpretive challenge in
    substructure-level attribution derived from fingerprint representations.}
    \label{fig:limitation}
\end{figure}

\subsection{Attribution Limitations from Fingerprint Overlap}

One observation worth noting is shown in Figure~\ref{fig:limitation}, where arrows
point to a methyl group adjacent to an unmethylated backbone nitrogen, both of
which appear in deep blue with nearly identical intensity. Chemically, a methyl
group would not be expected to contribute against membrane permeability, and
certainly not with the same strength as an NH group. This misattribution arises
from how Morgan fingerprints featurize local chemical environments. Because the
fingerprint encodes circular neighborhoods at increasing radii, a methyl group that
sits adjacent to an NH group will share overlapping environments at radius 1 and
radius 2. When the bit corresponding to that NH neighborhood activates and receives
a negative SHAP score, the equal division step distributes that score across all
atoms in the environment, including the neighboring methyl. The methyl therefore
inherits anti-permeability attribution, likely not because of its own chemical
character but because of its proximity to the NH group.

The contour-based heatmap
partially mitigates this by presenting the attribution as a continuous regional
signal rather than discrete per-atom values, encouraging the user to interpret the
highlighted region as a whole rather than individual atoms. A user with sufficient
biochemistry knowledge would recognize which feature in that region is the true
driver, but the potential for confusion remains, particularly for less experienced
users. This represents an inherent trade-off in fingerprint-based attribution.
It is also important to note that Morgan fingerprints encode molecular
topology rather than three-dimensional conformation, so no choice of radius
or bit-vector size can capture the conformational behavior known to
influence permeability in this molecule class, which we present as a use
case.

\subsection{Generalization Beyond the Training Distribution}

Figure~\ref{fig:ui} presents a more encouraging observation, which shows MolExplain
applied to a small molecule rather than a cyclic peptide. Despite the model being
trained exclusively on cyclic peptides from CycPeptMPDB, the attribution pattern
on this small molecule aligns well with general chemistry expectations. The
carboxylic acid and ester groups, both polar hydrogen bond donors or acceptors, are
highlighted in blue as anti-permeability contributors, while the benzene ring, a
nonpolar aromatic system, appears in red as a pro-permeability contributor. This
conforms with established physicochemical principles, suggesting that the Morgan
fingerprint representation captures substructural patterns general enough to
transfer beyond the training distribution.

Because Morgan fingerprints encode
circular chemical environments, associations learned between substructural patterns
and permeability in cyclic peptides appear to carry a meaningful signal for other
molecular classes. It is important to note that this is not a validated claim.
Properly establishing generalizability would require evaluation on a labeled
small-molecule dataset, but it does suggest this framework may extend beyond its
training domain with additional data.

\subsection{Future Work}

There are currently four concrete directions for
extending MolExplain.

\textbf{Reducing attribution overlap.} The misattribution described in Section~7.1
could be addressed by exploring different combinations of bit vector size and
fingerprint radius to reduce neighborhood overlap. Concretely, a thorough grid
search over radius and bit vector size, evaluated against known
permeability-active substructures in the test set, could identify a configuration
that balances substructural coverage with attribution resolution.

\textbf{Interactivity with the Output.} Adding an additional dimension of
interactivity to the system is a natural next step. The heatmap is currently
a static, server-rendered SVG, where red and blue regions convey which
substructures the model attributes to the prediction. A skeptical user may
also want a quantitative representation alongside the qualitative view the
heatmap offers. One possible implementation is a hover or click interaction
exposing the specific fingerprint bit(s) and atom radius contributing to a
given region's score, letting the user move from trusting the visualization
to inspecting it directly.

\textbf{Multi-property optimization.} In the use case presented, aggressive
N-methylation was pursued to maximize membrane permeability. But in practice, such
modifications could compromise the peptide's binding affinity to its intracellular
target, rendering its therapeutic potential irrelevant. This motivates a
multi-property display where the user can observe attribution panels for several
properties side by side and reason about how a structural modification shifts each
one simultaneously. However, this introduces an HCI design challenge: how many
properties can be displayed at once before the interface overwhelms the user? Too
many panels introduce cognitive overload, while too few may constrain the user's
ability to reason about meaningful tradeoffs. One candidate design is a scrollable
side-by-side layout capped at three or four properties, with visual indicators
flagging atoms where attribution directions conflict across properties.

\textbf{Session history.} A session history feature that allows users to track
their modification steps and download their final molecule would further support
the iterative design workflow, giving users a record of which structural changes
led to meaningful improvements. In practice, this could take the form of a
timestamped log that records each submitted structure alongside its prediction
result, with an export option that bundles the SMILES, predicted probabilities,
and heatmap images into a single downloadable file.

%-------------------------------------------------------------------
\section{Conclusion}

MolExplain demonstrates that combining fingerprint-based machine learning with
SHAP attribution and interactive visualization produces a tool that goes beyond
prediction to support hypothesis-driven structural exploration. Rather than
returning a bare probability score, the system exposes which substructures drive a
prediction and lets the chemist act on that information directly, closing the loop
between model output and structural insight. The use case presented is consistent
with established chemistry: MolExplain's attribution independently
recovers the known role of backbone N-methylation in improving cyclic peptide
membrane permeability, without that mechanism being encoded anywhere in the model
or the featurization. A further observation, that the same attribution pattern
aligns with general chemistry expectations when applied to a small molecule outside
the training distribution, suggests the underlying representation may capture
structural signal that generalizes beyond cyclic peptides, though this remains an
informal observation rather than a validated claim. Refining attribution precision,
assessing generalizability on labeled small-molecule data, extending the model
registry to additional properties, and validating attribution-guided modifications
against domain-expert review or wet-lab assays are the next steps for MolExplain as
a platform for explainability-driven molecular design.

\bibliographystyle{ieeetr}
\bibliography{MolExplain_citation}

\end{document}